\documentclass[runningheads]{llncs}
\usepackage{graphicx}
\begin{document}
\title{Snooping on Snoopers: Logging as a Security Response to Physical Attacks on Mobile Devices\thanks{This project was partially supported by FCT through LASIGE Research Unit funding, ref. UIDB/00408/2020,  LARSyS Research Unit funding, ref. UIDB/50009/2020, and project mIDR (AAC 02/SAICT/-2017, project 30347, cofunded by COMPETE/FEDER/FNR).}
}
\titlerunning{Snooping on Snoopers}
%
\author{Tiago Guerreiro\inst{1} \and
Ana Pires\inst{1} \and
Luís Carriço\inst{1}}
\authorrunning{Guerreiro, T. et al.}
%
\institute{LASIGE, Faculdade de Ciências da Universidade de Lisboa, Portugal}
\maketitle              
\begin{abstract}
When users leave their mobile devices unattended, or let others use it momentarily, they are susceptible to privacy breaches. Existing technological defenses, such as unlock authentication or account switching, have proven to be unpopular. We conducted interviews to uncover practices users currently engage in to cope with the threat, and found that it is common for users to try to keep their devices under close supervision at all times. One obstacle to this strategy is that displaying such protective behavior can be detrimental to social relationships. To address these concerns, we built a software tool that gathers activity logs in the background. Logs can later be reviewed as a time-line of opened apps and the actions performed within each, with events decorated with pictures captured inconspicuously with the front-facing camera. We evaluated this approach in a user study, and found participants to be generally eager to adopt the technology, although in different ways. Most users foresaw using it as a deterrent, or to check if they were snooped on, if that suspicion were ever to arise. Yet, some voiced the intention of creating “honey traps”. The results highlight both the opportunities and the potential dangers of the logging approach.

\end{abstract}

\section{Introduction}
Currently, users are susceptible to the threat of physical intrusion to their personal mobile devices, perpetrated by people known to them. This threat is neither remote nor improbable \cite{Marques16}. An increasing number of people have the most private aspects of their lives kept as records in small devices, whose security mechanisms are of dubious effectiveness against the adversaries most interested in their lives, and with most access \cite{Matthews2017,Marques19,Levy2020}.

To protect against physical intrusion, mobile devices typically employ authentication, both at the start of operation, and, in some cases, to access specific functionality. In practice, however, authentication is often not adequate to user needs, as is revealed by its underwhelming adoption \cite{Harbach2014,Egelman2014,Mahfouz2016}. The need to perform a cumbersome action at the beginning of every interaction causes many not to set it up. Additionally, with the most popular authentication methods, people in close proximity can easily observe the owner entering the code \cite{VonZezschwitz2013,vonZezschwitz2015_b,Wiese2016}. Finally, the security of unlock authentication is dependent on the owner not losing control over the device within a session, an assumption that is increasingly not reasonable in the case of known non-owners \cite{Hang2012}. When the owner hands over the device for someone to place a call, that person can easily look through the call log. When the owner shares it to show a picture, sensitive content may be a couple of swipes away.
This kind of \textit{social sharing} also makes multi-user support, another well-known security mechanism, ineffective. Commercial mobile devices have introduced features like user accounts and profile restrictions, often with the intent of preventing misuse by children. They, however, require configuration, and an explicit transition action, which is at least as cumbersome as authentication.
For end-users the choice then seems to be between either to absolutely trust, or to absolutely distrust, people they know -- a proposition that falls in stark contrast with people's desire for granular regulation of privacy \cite{Palen2003,Matthews2016}.

This research aims to bridge this gap with security capabilities that accommodate the social relationship between self and others. In particular, we evaluate the acceptability of technologies that continuously monitor the possibility of misuse, and gather information that can assist in lessening its effects – capabilities which have been typically associated with intrusion detection and response systems in networked computers.

In this paper, we first report on a formative study with smartphone users. In 15 interviews, we inquired about user concerns over physical intrusion by their close-ones, what they do to avoid it, and the adequacy of those strategies from their perspective. We found that people worry about the possibility of intrusion when they leave the device unattended, and when they share it momentarily. To cope with the threat, many people resort to counter-measures they can enact on their own, instead of any particular security technology. Examples include self-censorship of content retained on the device, and defensive social practices, such as never letting the device out of sight, often in a balancing act between protecting oneself and not displaying mistrust of others. 

\begin{figure}
    \centering
    \includegraphics[width=0.75\textwidth]{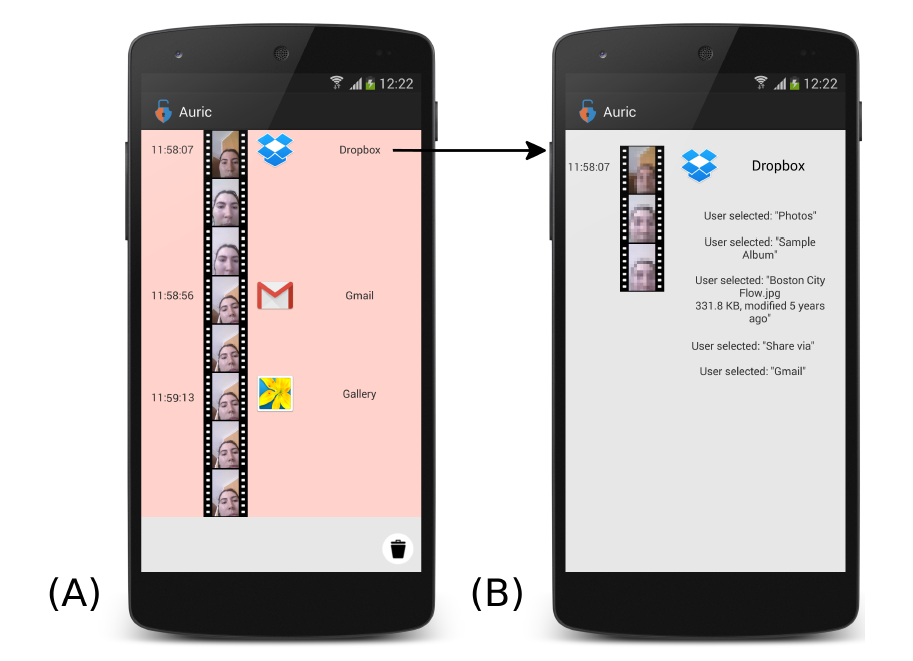}
    \caption{Visualization of logged activity on a smartphone. \texttt{(A)} shows a time-line of apps used in a session, decorated with photos taken with the front-facing camera. \texttt{(B)} shows user actions within an app.}
    \label{logs}
\end{figure}

To deal with both the use case of unattended access and the one of social sharing abuse, while addressing users concerns over being perceived as overly protective, we then explore the feasibility of activity logging as a security response. Activity logging offers the user the opportunity to know if the device was snooped on, without having to engage in any explicit action, such as changing users accounts. It can also act as a deterrent: if potential snoops know that the device records activity, they may reasonably fear being detected. To test the practicality of this concept, we built a logging tool with information gathering capabilities for Android. The software runs on the background as the device is being used, collecting logs of application launches, user actions within those applications, and pictures taken with the front-facing camera. The logs can be reviewed in time-line format, decorated with the pictures (figure \ref{logs}).

In a second study, using the developed tool, we evaluated the appropriateness of this approach, and in particular how usable logging could be adopted by end-users. Most participants were keen on having the technology available to them, although they were ambivalent on exactly how they'd put it to use. Aside from the practical aspect of being able to verify if intrusions had occurred, we expected most participants to see most value in deterrence. What we found, however, is that there's often a stronger desire to know who, in one's social circle, would breach the trust in them deposited, even at the cost of having private information accessed. Some users went as far as expressing the intention of using the logging capabilities offensively, actively creating situations where others could be tempted to use their devices without permission.

In summary, the contributions of this paper are:
\begin{enumerate}
\item It shows how user concerns over physical intrusion by socially-close adversaries remain unresolved, as is revealed by existing coping mechanisms.
\item It shows the feasibility of providing security logging capabilities in a way that is suitable for end-users.
\item It explores the road to adoption of such approach, looking at both its suitability to address user concerns, and to possible dangers.
\end{enumerate}
\section{Related work}

\subsection{The threat of physical intrusion}

Personal mobile devices frequently contain sensitive information about the owner, including access codes, personal communication, call and text logs, contacts, pictures, videos, and location records \cite{Ben-Asher2011,Felt2012,Muslukhov2012}. Users are acutely aware and concerned about privacy risks relating to their smart devices \cite{Chin2012,Clarke2005,Karlson2009}. Indeed, those concerns aren't unreasonable: in the last few years, while adoption of personal mobile devices has rapidly grown, so the risks have materialized, for instance in the form of mobile malware \cite{Felt2012}.

Research on the security challenges raised by modern personal mobile devices has focused on scalable technological attacks, like malware, that traditionally have been seen as the most damaging. Recently, as spearfishing and the insider threat have gained more attention in network security, so have socially close-adversaries been recognized as a threat to personal mobile computing \cite{Muslukhov2013,Levy2020,Marques16}. In reality, for the end-user, security incidents from malware threats are rare, and, when effective, tend to have low impact, mainly in the form of unsolicited advertising \cite{Felt2012}. Concerns of younger users, the so-called "digital natives" align with this reality: they are more aware of threats with a social context (loss, theft, snooping, shoulder-surfing) than sophisticated remote attacks \cite{Kurkovsky2010}.

Physical intrusion by socially-close adversaries is frequent. In a survey of internet users, 14\% of participants reported that "someone used my mobile phone without my permission with intention to look at some of my data", and 9\% that "I used someone’s mobile phone without owner’s permission to look into their data". \cite{Muslukhov2013}. In a follow-up study, \cite{Marques16} through an anonymity-preserving online survey found that prevalence was even, particularly among younger generations.  In a national survey of the US population \cite{Boyles2012}, 15\% of smartphone owners reported having had "another person access the contents of their phone in a way that made them feel their privacy was invaded". Since it is not always possible to know that one's phone was accessed without permission, it follows that the actual prevalence of physical intrusion should be even higher. It is also on the rise: in the 18-24 age group, 24\% say that someone else has accessed their phone in a way that compromised their privacy.
In Symantec's Honey Stick Project, it was found that 9 out of 10 strangers, when faced with the opportunity to access sensitive data on a “lost” phone, do so. Unsurprisingly, then, the main security concern of users is that someone manages to obtain their devices \cite{Chin2012}. In another recent study \cite{Hayashi2012}, when presented with the possibility of being able to prevent socially-close individuals from accessing some functionality on their phone, 7 out of 10 participants reported they would like to do so. But perhaps the most revealing trend is the booming of a spyware industry, catering to cravings for information contained in mobile devices of socially-close individuals \cite{Marques19}.

\subsection{Defenses}

Currently, the legitimate concerns over physical intrusion by socially-close adversaries are addressed, in research and industry, mainly by authentication solutions. Recent years have seen considerable efforts in improving both the security and the usability of mobile authentication methods.

An issue often associated with common authentication methods, like passwords, PINs and graphical passwords, is that they are susceptible to rudimentary attacks. Because mobile devices are often used in uncontrolled surroundings, an attacker can observe the secret code as it is entered \cite{Schaub2012}. Socially-close adversaries, in particular, have many opportunities to shoulder-surf. Even if not observable in real-time, in touchscreen devices it is often possible to at least narrow down the authentication code from smudges left in the surface \cite{Aviv2010}. Attempts to make mobile authentication non-observable are too many to enumerate, but, more often than not, the added security comes at the expense of a greater burden on the end-user's attention. Yet, many users already do not set authentication up on their personal devices because it requires too much effort \cite{Egelman2014,Harbach2014,VonZezschwitz2013,vonZezschwitz2015_b}.

Recently, smartphones have started to provide users the ability to authenticate with \textit{something they are}, through biometric unlock methods. Devices offer fingerprint authentication, as well as facial recognition (Face Unlock), and, more recently, voice recognition (Trusted Voice). These biometric methods have fallbacks to secret-based methods. For instance, in TouchID, users must authenticate with a passcode at every reboot. Even if they never reboot, passcodes are required periodically, presumably to avoid users forgetting them. Empirical studies indicate that TouchID users do not choose better passcodes for fallback authentication \cite{Cherapau2015}. Biometric unlocking is thus subject to many of the same security limitations that apply to secret-based unlocking. 

Recognizing that usability is a key factor in adoption, another approach that has emerged is to reduce the frequency in which explicit authentication is required. To achieve that, flavors of continuous background authentication, in which characteristics of the operator's interactions are taken as a behavioral biometric, have been suggested \cite{Kayack2014}. Some proposals use contextual information, such as geographical location, as an additional authentication factor \cite{Hayashi2013}. 
Attempts to make authentication more usable, or more secure, or both, have been made. Despite these advances, little consideration has been given to security mechanisms aside from authentication. As far as we known, this paper is the first to explore the feasibility of user-facing logging tools as a deterrent to known non-owner intrusions to modern personal mobile devices.

\section{Study 1: Privacy-related practices}

We conducted a formative interview study (n=15) to understand user concerns about physical intrusion and their current defensive strategies. Our objective was to capture a broad range of observations from a diverse pool of participants, and not to quantify their prevalence. Hence, we opted for a semi-structured format, restricting the questions to the same three general themes: concerns (things people are worried about), practices (things people do or don't do to safeguard), and incidents (specific episodes of negative experiences), all regarding possible intrusion by people in close social circles. We conducted the interviews remotely, in an online instant messaging platform. The social distance and informality of online communication, we reasoned, would make it easier for participants to share negative experiences.

\subsection{Participants}

We recruited participants from an online community of Amazon Mechanical Turk workers. Participants were invited by personal message by a community manager. The advertisement targeted only people who regularly use smartphones. We elected to recruit from this forum because of the known diversity of Mechanical Turk workers \cite{Buhrmester2011}. Participants were invited in small groups of varying sizes, until the interviewer was confident that no new information was being produced. They were rewarded a \$7 gift card. Table~\ref{users1} lists the characteristics of participants in this study.

\begin{table}[tbh]
\begin{tabular}{lllllll}
\hline
\textbf{} & \textbf{Gender} & \textbf{Age} & \textbf{Country} & \textbf{Lives With} & \textbf{Mobile Devices} & \textbf{Locks devices} \\ \hline
P1        & Male            & 18-24        & USA              & Family              & Smartphone/Tablet       & Password/Password      \\ 
P2        & Female          & 18-24        & USA              & S.O.                & Smartphone              & Password               \\ 
P3        & Female          & 25-34        & USA              & Alone               & Smartphone/iPod         & Password/No            \\ 
P4        & Male            & 25-34        & India            & Family              & Smartphone/Tablet       & Fingerprint/No         \\ 
P5        & Male            & 25-34        & USA              & Family              & Smartphone              & Fingerprint            \\
P6        & Male            & 25-34        & USA              & Family              & Smartphone/Tablet       & PIN/PIN                \\ 
P7        & Male            & 35-44        & Canada           & Family              & Smartphone/Tablet       & No/No                  \\ 
P8        & Female          & 35-44        & USA              & Family              & Smartphone/iPod         & No/No                  \\
P9        & Male            & 45-54        & USA              & Family              & Smartphone/Tablet       & No/No                  \\ 
P10       & Female          & 45-54        & USA              & Alone               & Smartphone              & Password               \\ 
P11       & Male            & 45-54        & USA              & Family              & Smartphone/Tablet       & Password/Pattern       \\ 
P12       & Male            & 45-54        & USA              & Roommates           & iPod                    & No                     \\ 
P13       & Female          & 55-64        & USA              & Family              & Smartphone/Tablet       & No                     \\ 
P14       & Female          & 55-64        & USA              & Alone               & Smartphone              & No                     \\ 
P15       & Female          & 55-64        & USA              & Family              & Smartphone/Tablet       & No/No                  \\ \hline
\end{tabular}
\label{users1}
\caption{Participants in the reported exploratory interview study, about concerns over physical intrusion and defensive practices. A remote and diverse pool was recruited, in order to capture a breadth of experiences.}
\end{table}

\subsection{Procedure}

Participants were sent a link to a private chat room, and given instructions on how to initiate a conversation with the interviewer. In the beginning of the interview, we explained the purpose of the study and the procedure, and asked for consent to proceed. The interviews had a 30 minute time limit. They started with some demographics and general characterization questions; then moved to general questions about user concerns, which, through probing and eliciting recent examples, would push the discussion towards continued practices, and, when that was the case, to specific negative experiences.

\subsection{Analysis}

Given the exploratory nature of study, we opted for an inductive approach, where we try to identify the most important types of participant experiences, and illustrate them with first-person accounts. The transcripts of the interviews were coded to assist in that process, and to avoid an analysis skewed towards pre-conceived notions. The researcher that conducted the interviews compiled a set of codes by analyzing the first 8 interviews. Two other researchers then both coded a subset of 5 interviews with this book. Reliability was at an acceptable level for this subset (average Cohen's $\kappa$ = 0.85). Hence, one of the researchers coded the remaining 10 interviews.

\subsection{Findings}

\subsubsection{Concerns}

We identified general concerns users have about their personal mobile devices, as they relate to intrusion by people in their social circles. 

Concerns with known non-owners are often dictated by usage. Some people don't use their devices in such a way that it would keep data they would consider private. For those people, there are still worries about \textbf{misuse}:

\begin{quote}“I would be worried that they [family members] would do something to mess it up, but there is really nothing that I would have a problem with anyone seeing. I do not handle any finances, do any work or anything really important on it.” (P9)\end{quote}

\begin{quote}“I have about five games that I am in the middle of.  I have them in a separate folder and told them [other people that sometimes use the device] not to play those games.” (P12)\end{quote}

Concerns over \textbf{unauthorized access to sensitive information} come in several varieties. While some participants report not being concerned over their own devices being snooped on, they projected the threat onto their close ones, as, for instance, P9:

\begin{quote}“For my kids, I would not want people having access to their websites, photos and other data they have on their phones.”\end{quote}

Worries about sensitive content are, even among people in close social circles, highly \textbf{person-dependent}. There isn't a direct relationship between social distance and level of concern. Often, it is the opposite: the closer a person is, the more concerned participants were. P4, for instance, reported:

\begin{quote}“If a family member or a kid had taken my phone, I would feel more concerned. If it's a friend, then not so much.”\end{quote}

For people that see their devices as highly personal, we observed a greater tendency to be concerned over privacy invasion, even when the device is only momentarily handled by others. There is a kind of \textbf{separation anxiety}, even if no particular threat can be articulated. Some participants reported emotional distress in these types of situations:

\begin{quote}“Yeah, I always get uneasy when someone has my phone even for a little bit and almost sort of start to grab back for it.” (P2)\end{quote}

\begin{quote}“[when someone else is using my device I feel] protective, maybe a little tiny bit nervous. Even if you trust the person not to, you can't control whether or not they go poking around in some other app or something.” (P3)\end{quote}

Despite the anxiety it causes, people still let others handle their devices. Unspoken social norms sometimes mandate that people put themselves in vulnerable situations. P10 articulated this point:

\begin{quote}“I've given my phone to people to use. [...] I can't say no if someone needs to make a call. It feels wrong.”\end{quote}

Accounts of user concerns over close-ones accessing private data kept on mobile devices have previously been reported \cite{Karlson2009,Hang2012,Marques19,Levy2020}. The testimonials we have gathered further lend support to the notion that displaying excessive control or denying access causes a great deal of unease. There remains, it appears, a conflict between the desire for control, and what is socially-acceptable behavior.

\subsubsection{Practices}
We also sought to understand what people did to protect themselves against intrusion. If the concerns that ours and previous work identified subsist, it stands to reason that people somehow must have found strategies to cope with them, even if they are not optimal.

In some cases, people invoke that they don't do anything in particular to protect themselves, because \textbf{no one would be interested} in what they keep on their devices. This is expressed, for instance, by P15:

\begin{quote}“My sister asked me to delete some texts recently from a conversation we had about my son. It was kind of insensitive [about a personal issue]. I don’t think he would snoop anyway. Yes, I don’t think that they [the children] think they would find anything interesting.”\end{quote}

Others invoke \textbf{lack of ability of other people} that could have physical access. Some potential intruders aren't a threat simply because they wouldn't know how to use the device. This is often the case with children and older family members. P8 exemplifies this rationale:

\begin{quote}“They [the children] like their own games and shows so would not "wander" into anything else. Plus, they would have a hard time understanding other things.” (P8)\end{quote}

Often people combine lack of ability and interest of others with a \textbf{reliance on trust relationships} as reasons to not take protective measures. Different kinds of informal access controls apply to different people. Examples from participants include:

\begin{quote}“I trust all of those who are close to me. My wife and children would not make purchases. They have their own phones as well. I trust they have no interest in accessing my banking, or online accounts. I trust that my friends and coworkers have no interest or desire to access my device. I realize that is naive, but I can't imagine a co-worker picking up my device.” (P7)\end{quote}

\begin{quote}“I don't snoop, so they don't either. My brother would ask before he looked at anything else on the phone. My mother wouldn't know how to snoop on the phone – [but] she would maybe try.” (P10)\end{quote}

Participants also report that not only the type of relationship, but also the \textbf{kind of device}, is important establishing socially-imposed access policies. Generally, the smartphone is seen as highly personal, in opposition to other mobile devices:

\begin{quote}“Well, my phone is usually on me, but my tablets are just laying around in the house. My wife does not have to ask [for permission to use them], but kids will.” (P9)\end{quote}

\begin{quote}“I just don't let my cell phone out of my sight / hands / purse when I'm outside my house. My iPod never leaves the house but I also do not save passwords on any personal apps so I have to log in each time on those.” (P8)\end{quote}

As suggested in the previous examples, one common measure against the threat of someone using the device without permission is \textbf{keeping it close}. This is often combined with reliance on trust relationships: 

\begin{quote}“When I am outside of my home my devices are always in my pockets or in my view. When I am at home I don’t get too far from them but they have lied around. I am not too concerned with the people in my home causing me trouble.” (P6)\end{quote}

\begin{quote}“I trust the people at work and at home and the phone is locked, and if I'm out somewhere the phone is most likely in my pocket or in my purse.” (P2)\end{quote}

For some, however, trust relationships seem insufficient. Even surrounded by trusted people or in a trusted environment, some participants reported efforts to keep the devices under control: 

\begin{quote}“I never leave them [other people] alone with the phone.” (P5)\end{quote}

\begin{quote}“When I take my girls to their gymnastics class, I will leave my purse and ask a friend to keep an eye on it. [I] will leave my wallet but will not leave my phone.” (P8)\end{quote}

A common practice today, and something that most participants reported, is showing other people content on the device, like pictures. In these situations, a common defense against the threat of intrusion is \textbf{not letting others hold the device}. This was expressed by several interviewees:

\begin{quote}“If I show a picture it's not handed over, and no-one but me or my wife uses my phone/tablet. Looking don't require it to be held.” (P11)\end{quote}

\begin{quote}“If I want to show someone something (a photo, for example), I hold the phone where they can see it.” (P14)\end{quote}

Finally, another common practice is \textbf{self-censorship}. 
Many participants reported not keeping private information on the devices as a preventative practice:

\begin{quote}“I also think we have to accept some responsibility for what is on our phones and devices. I try to be careful about what I keep on all my devices, even my Kindle.” (P10)\end{quote}

\begin{quote}“I also don't put anything in a text or online that I'm unwilling to have become public.” (P14)\end{quote}

While most reports of self-censorship were alluded to being preemptive, one participant reported to actively clean-up whatever is deemed sensitive.

\begin{quote}“I usually delete my texts after I read them, and for any pictures that I do not want to share with anyone, I delete them after saving it in my Dropbox.” (P4)\end{quote}

Lacking the support in either social norms or in technology, participants reported behaviors that may well be described as coping strategies. Believing that close-ones aren't able or interested in private contents doesn't seem as much a carefully considered choice, as it does seem to be a way to put the problem out of mind. Engaging in aggressive protective behaviors and self-censorship is indicative of the extent to which the technological defenses aren't adequate. In this, we find parallel with established literature on password habits, which shows that when the technologies are at odds with user needs, people find a way to lessen the burden for themselves, at the expense of security \cite{Adams1999,Koushki20}.

\subsubsection{Negative experiences}
We collected a number of negative experiences that, taken together, suggest that common defensive practices aren't always effective, as they collide with social norms. 

For instance, P10 reported that, despite keeping the device close, people have used it unexpectedly, in circumstances where an \textbf{overt negative reaction could be deemed as antisocial}: 

\begin{quote}“At work during lunch I often have my phone on the table. People have walked by and said 'oh is that the 5C or 5 or is that the 6?' [then] grab and swipe. I just would never think to pick up something so personal. People have emails and photos and texts. It is really uncomfortable.” (P10)\end{quote}

This social tension is evident in other reports, where, despite social pressure, participants reported actually taking steps to regain access to the device when others were using it, even they were not accessing sensitive content:

\begin{quote}“There have been times when people were 'just looking' at it and it made me anxious.  So I have jokingly wrestled it back from them! I'm sure I came across as pathetic, but it isn't just a phone anymore. If someone has your phone they now have your email, your photo album, your banking info, your apps, your recent purchases, books you've downloaded, videos you've watched. Not just a phone. I'm not paranoid, honest!!” (P10)\end{quote}

\begin{quote}“After a staff meeting at work, I had an image to share with a colleague. He had a laugh, and others wanted to see so it got passed around the table. About 3/4 of the way around, I announced: okay, gimme my phone back. There was no inappropriate or embarrassing content on my phone. In retrospect, must be due to the possibility that my wife might send a racy or inappropriate text message.” (P7)\end{quote}

We also collected two reports where participants had to take steps to thwart situations that were escalating towards privacy invasion, and even so had to do it jokingly: 

\begin{quote}“I was talking [messaging] to my ex-girlfriend and she [my cousin] took my phone and I felt uncomfortable as I didn’t want to share the conversation with anyone else. I was messaging with her. She [the cousin] snatched it from me as joke [but I got it back soon after]” (P4)\end{quote}

\begin{quote}“One of my brothers actually opened up my texts.  He quickly shut the app though when I asked what he was doing. I [...] made him choke on his drink by suggesting that if he kept going he would be seeing some naughty photos. None there, just a threat, [but] no brother wants [to see] that!” (P10)\end{quote}

Although it is certainly not always the case that recovering control is done displaying playfulness, we, understandably, did not receive such reports. However, the fact that in the reports we collected participants acted light-hardheartedly, so as not to damage existing social relationships, highlights the fact that current defenses aren't, at times, appropriate.

\section{Building a mobile activity logger}
To explore the feasibility of the activity logging approach, we developed a tool for Android, designed to be accessible to end-users. The software captures user behaviors, in the form of interactions with applications throughout time, taking advantage of the "app model". Since functionality is distributed across small software programs, when someone uses a specific app, the purpose is often clear. Moreover, the actions performed within each app are captured using accessibility hooks in the operating system, without breaking its security model. The logs are shown side-by-side with pictures taken with the front-facing camera, for context.

\subsection{Concept}
This approach is grounded on security technologies in non-consumer computing assets. Large computer networks have for long maintained intrusion prevention systems (IPS) as a second-line capability, aimed at detecting security threats that have found a way into the organizational perimeter despite primary security barriers, and mitigate them, either through counter-actions, or by gathering and keeping information that can help in incident recovery \cite{Scarfone2007}. In recent years, a great deal of consideration has been given to the insider threat – the risk that an individual inside the organization conducts an attack. The approach we propose parallels such efforts in the realm of personal computing devices, where the insiders are people known to the owner.

Activity logging is a type of response that host-based IPS can provide to security threats. For the purposes of this research, we limit the concept to a service that runs on the background, as to not interfere with regular operation, and collects time-stamped information about which applications were interacted with, what actions were performed in those application. To discern the source of those interactions, the logger also collects pictures taken with the front-facing camera. Such security mechanism can run concurrently to other deterrents to physical intrusion, like unlock authentication, if the user so chooses. 

Unlike the typical IPS, a major design consideration for a personal activity logger is the ease of operation. In an organization, when a security incident occurs, specialists conduct analysis of security logs using complex tools and techniques. For end-users, the logs must be made available in usable fashion. In the tool we built, we decided to address this requirement using a time-line of actions, and a film roll analogy for displaying the pictures. To facilitate navigation in logs, we also added the possibility of users filtering out sessions where no other faces other than theirs were detected. We note that despite the known problems with accuracy in face matching techniques, false positives and false negatives do not impose severe security or usability problems in this implementation. A false negative does not mean that an intrusion is not recorded, only that it is not listed when the user applies the filter. A false positive means that the user will perhaps look through more logs can it optimally could. We further note that, since many pictures are taken in each session, there is opportunity to improve accuracy with repeated classification.

Modern mobile devices, like smartphones and tablets, have characteristics that align with the typical capabilities of an IPS, namely for \textit{logging} and \textit{information gathering} \cite{Scarfone2007}. A logger in a general-purpose IPS monitors only the effects of suspicious activity, such as network traffic, file and operating system calls, and changes in configuration. However, in a personal mobile device, an attack can be perpetrated by just looking at the content. Mobile devices offer capabilities to record the activity itself. Although, system-wide monitoring is usually restricted according to the operating system's security policies, in Android it is possible to obtain a representation of user interface activity using accessibility hooks. The sensing capabilities of smartphones and tablets also allow additional information gathering, with cues on operating context. Since we focus on physical intrusion by known non-owners, our implementation only captures pictures taken with the front-facing camera. Although it would be possible to collect, for instance, geographical location, this kind of information isn't particularly relevant, as it would be if the threat under consideration was loss or theft.

\subsection{Implementation}
To validate our proposed approach, we built an AURIC prototype for Android. We constructed a service that runs in the background, gathering activity logs and additional information; and an application, which offers an UI for auditing the records collected by the service.


\subsubsection{Logging}
The logging component collects data for every session in its entirety, starting when the user goes beyond the lock screen, and until the screen is powered down. It uses Android's Accessibility Services API, which fires events when something notable happens in the user interface. A custom-built semantic filter translates the raw stream into human-readable units. For instance, when a user writes in a text field, accessibility events are fired for every key press, and the filter merges the input in a single string.
Using the Accessibility Services API in security applications is not unheard of. Many password managers use this functionality to access fields inside running apps.

We initially developed the ability to collect logs as video replays, by intercepting the low-level system-wide GUI video stream. To do this, however, the device has to be rooted. We didn't expect this option to be currently viable for most users, as it requires both advanced knowledge and the breaking of Android's security model. Hence, the prototype defaults to collecting logs through the accessibility hooks, although the ability to collect video streams remains as an option for advanced users.

\subsubsection{Information gathering}
The information gathering component uses Android's Camera API to collect samples periodically using the front-facing camera, in configurable intervals. To flag pictures as being of the owner or not, for filtering purposes, our implementation uses OpenCV as a classifier. At runtime, OpenCV is first used to detect if the picture contains a face or not. If it does, the sample is matched against three portraits of the owner collected at enrollment. The best similarity score is recorded as metadata, alongside with the picture itself, in the database. 

\subsubsection{User interface}
A logging tool is only effective when the owner can discern potential security incidents. We therefore designed a user interface that facilitates accessing the logs that were collected. At the top level, our prototype shows a calendar view that marks the days for which logs were collected. The user can apply a filter, which colors the days when there were sessions in which a face other than the owner's was detected (according to a configurable similarity threshold). When a day is selected, a list of sessions is shown, marked with the start time, and colored red if the filter was applied. 

Upon selecting a session, a list of apps that was opened is shown, alongside a camera roll of pictures taken with the front-facing camera (figure \ref{logs}A). Upon selecting an app, a second log screen is shown, with the actions that were performed while using that app (figure \ref{logs}B).
At runtime, the service can optionally show notifications that it is running on the top bar and over the lock screen. The notifications alerts the operator that the device has the ability to record user actions for later review.

\subsubsection{Limitations}
A technical consideration that should not be overlooked when evaluating technology for mobile devices is resource consumption. Informal testing on a Samsung S4, a lower bound, indicated that in a period of 8 hours of regular operation, the logger was responsible for 5\% of battery usage, and occupied 140 MB of data storage (which could conceivably be offloaded to the cloud). We did not perform any formal testing, nor did we optimize the prototype to minimize consumption, as our purpose at this point was to demonstrate the feasibility of a solution of this kind to be made available to the general public.

\section{Study 2: adoption}
We conducted a second interview study (n=12) to understand how user-facing logging tools might be adopted. Using the developed tool as a design probe, we conducted semi-structured interviews in a lab environment, with questions regarding participants’ willingness to adopt, projected ways of using them in their daily lives, and the implications in their social relationships. We opted for a semi-structured and conversational format, which would favor a common understanding of the discussion as being about the general implications of logging tools, instead of a particular embodiment.

Although perhaps more could have been uncovered in a field study, where participants would use the prototype for an extended period of time, we judged the risk to subjects to be unacceptably high. If participants were to identify physical intrusions, they might have negative experiences that we couldn't mitigate. Moreover, we wouldn't have been able to prevent participants from recording the actions of people who did not consent to participate. 

\subsection{Participants}
We initially recruited a convenience student sample of smartphone users. We added participants until informally observing that new laboratory sessions were not adding new information. Having reached that point, we decided to recruit a small number of non-students, to see if new insights arose. Table~\ref{users2} lists the participants. The process took one week and participants weren’t compensated for their time. Interview sessions took 30 minutes to an hour.  

\begin{table}[tbh]
\centering
\begin{tabular}{llllllll}
\hline
\textbf{} & \textbf{Gender} & \textbf{Age} & \textbf{Occupation}    & \textbf{Lives with} & \textbf{Mobile devices} & \textbf{User} & \textbf{Locks} \\ \hline
P1        & Female          & 18-24        & Student                & Family              & Smartphone/Tablet       & Heavy                 & Pattern/Pattern        \\ 
P2        & Male            & 18-24        & Student                & Family              & Smartphone              & Light                 & Pattern                \\ 
P3        & Male            & 18-24        & Student                & Family              & Smartphone              & Heavy                 & PIN                    \\ 
P4        & Female          & 25-34        & Student                & Dormitory           & Smartphone              & Heavy                 & PIN                    \\ 
P5        & Male            & 25-34        & Student                & Roommate            & Smartphone              & Heavy                 & No                     \\ 
P6        & Male            & 25-34        & Student                & Family              & Smartphone              & Heavy                 & No                     \\ 
P7        & Male            & 25-34        & Student                & Roommate            & Smartphone/Tablet       & Heavy                 & PIN/PIN                \\ 
P8        & Female          & 25-34        & Student                & Family              & Smartphone/Tablet       & Heavy                 & No/No                  \\ 
P9        & Male            & 25-34        & Student                & Family              & Smartphone/HGC          & Light                 & No/No                  \\ 
P10       & Female          & 35-44        & Marketing & Family              & Smartphone/Tablet       & Heavy                 & PIN/Pattern            \\ 
P11       & Male            & 35-44        & Marketing & Family              & Smartphone/Tablet       & Heavy                 & No/No                  \\ 
P12       & Male            & 45-54        & Marketing & Family              & Smartphone/Tablet       & Heavy                 & No/PIN                 \\ \hline
\end{tabular}
\label{users2}
\caption{Participants in the reported evaluation interview study. To complement the convenience student sample, a group of non-students was recruited.}
\end{table}

\subsection{Procedure}
In the beginning of each session, participants were explained the purpose of the study and the procedure, and asked for consent to proceed. The session was structured in three parts. 

Part 1 was a brief interview that included demographics and a small set of questions aimed at understanding if participants had concerns when sharing devices or leaving them unattended. 

Part 2 was a dramatization of the two situations, using the prototype as a design probe. After a brief explanation, participants were handed a smartphone with the logging tool installed, asked to act as if it were theirs, and prompted to follow a wizard to enroll to face recognition. They were given a tour of some of the features, most prominently, the option to hide the notifications. The dramatization of the scenarios then started. The first was concerned with leaving the phone unattended. Participants were told to lock the phone, leave it on the table, and leave the room momentarily. The moderator then took hold of the phone, unlocked it, browsed through pre-seeded text messages, emails, and browser history, and put it back. Participants were called back and prompted to see the logs. The second scenario was concerned with abuse of social sharing. Participants were asked to open a picture and show it to the one moderator, handing over the device. A second moderator then continued to speak with the participant to simulate distraction. Meanwhile, the first moderator flipped to another picture, emailed it, and returned the phone in the state it was shared originally. Participant were then prompted to again see the logs.

Part 3 was an exit interview. Participants were first asked for first reactions, and then about how they would use a logging tool it in their daily lives, if at all. Throughout the interview, participants were asked to recall recent experiences, and to relate them to how logging could have change them.

\subsection{Analysis}
Since we had a mostly uniform sample and the interviews were focused on adoption issues, we opted for a mixed approach, where we quantify the qualitative data, and illustrate it with examples.
Audio of the interviews was recorded and transcribed for analysis. First, two researchers independently developed code books based on two interview transcripts each. Then, they met and agreed on a preliminary set of codes. Afterward, the same two researches re-coded the 4 interviews, compared the results, and, by consensus, agreed on an extended set of 47 codes. The two researchers then coded all the remaining interviews concurrently. Reliability was found to be acceptable (Cohen’s $\kappa$ = .92). For the remaining cases where there was disagreement, codes were omitted from analysis.

\subsection{Findings}
\begin{figure}[tbh]
    \centering
    \includegraphics[width=0.7\textwidth]{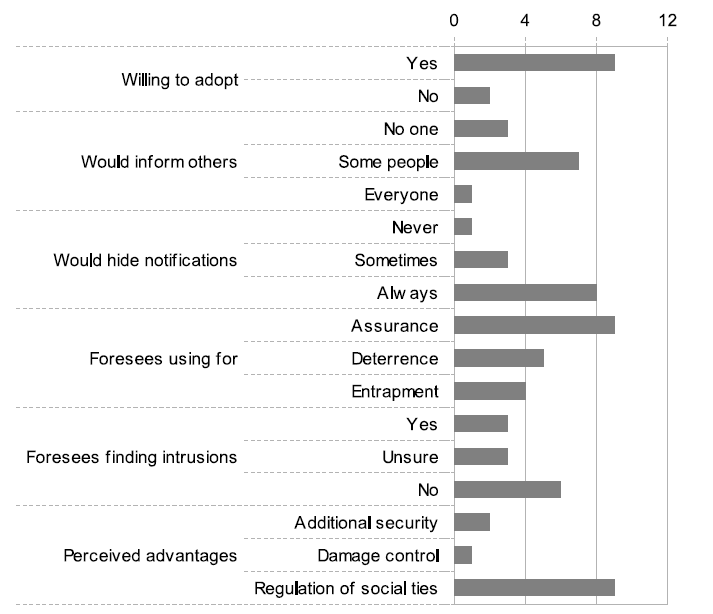}
    \caption{Frequency of the adoption-related codes in the second user study.}
    \label{results2}
\end{figure}

For brevity, we do not report on the findings regarding user concerns and practices, captured in the first set of questions, as they do not add new information to what was reported in the first user study. We also do not report on aspects of user experience, as they relate to the particular prototype we presented to users, and are uninformative to the larger community. We did find some minor usability issues, mostly with labels that users had difficulty interpreting. All users where able to identify the abusive activity in the logs, for both scenarios.
We instead focus on the results that inform on adoption issues. They are summarized in figure \ref{results2}.

\subsubsection{Willingness to adopt} 
In the exit interview, we asked participants whether they would like to install the application we demonstrated on their own devices (although that wasn't a real possibility). Ten participants showed interest. Some indicated their interest even before being asked specifically. For instance, P10, while seeing the logs in the first dramatization, said: 

\begin{quote}
"I'll take it."
\end{quote}

The two that weren't interested indicated that at least part of the reason was technical difficulties: one didn't have a front-facing camera, and the other didn't have enough storage.

\subsubsection{Letting others know} 
We probed participants as to whether they would announce to people close to them that they had an app that logged activity. We reasoned that if people intended to use a logger as a deterrent, letting others know would be a common strategy, as would be displaying notifications. 

However, only one indicated the intention of telling everyone about the logging capabilities, for reasons that go beyond deterrence:

\begin{quote}
"Yes [I would tell people]. This way they would know better than to mess with it; they know I'd find out. Also, because apps and gadgets are always conversations topics, and I like talking with my friends about it." (P12)
\end{quote}

Three indicated that they wouldn't tell anyone, and the remaining that they would announce it only to some people. We again observed that social distance is not a good proxy for trust; some participants reported that they wouldn't announce the capabilities exactly to those closest to them: 

\begin{quote}
"I think that some people would go without knowing. I think I wouldn't tell to those closest to me, just to see if the trust I have in them is justified or not." (P11)
\end{quote}

We also asked participants if they would have the service show or hide notifications, which would alert the operator that activity was being recorded, thus acting as a deterrent. Eight responded that they wouldn't, thus indicating that users prefer to closely regulate who knows about the logging capability and who doesn't.

\subsubsection{Projected usage} 
From the interviews, what transpires is that participants see three ways of using the tool. The majority of participants (9/12) indicated that they would use it for assurance purposes, that is, to have a way of knowing if someone ever did break in. Contrary to our initial expectation, only five of the participants expressed the intention of using as a deterrent, which explains why so few said they would announce it to others and want notifications to be shown.  Disconcertingly, four of them openly admitted to the intention of creating situations that would propitiate intrusion.

Three participants were indeed convinced that they would catch someone using their device without permission. Three were unsure, and the remaining six were certain that they wouldn't. The fact that although a majority of participants does not think that they would be a target of an attack, but still want the ability to know if they were, is, we believe, revealing of currently unmet needs.

\subsubsection{Usefulness} 
Seldom did participants identify what are commonly seen as advantages of intrusion prevention systems. Two offered general comments about the tool providing an added layer of security, which they considered useful. Only one participant referred to the advantage of allowing damage control.

In contrast, all but three participant indicated, has an advantage of logging over other security features, that it could be used to regulate social relationships, by finding the enemy within, and then adjusting the relationship. Comments about revealing the true nature of relationships abounded, and were always present when there was reference to adopting the tool for entrapment:

\begin{quote}
"We should live surrounded by people that we trust and wouldn't do such a thing. This tool would help in identifying those people. To have control over who you let in your life and consider to be a friend." (P5)
\end{quote}

\begin{quote}
"Those people will access critical data, but at least I'll know. That dude, I'll never trust him again." (P9)
\end{quote}

We found a worrying trend. Despite participants wanting the tool, as it addresses currently unmet needs, there are social implications of making it available that cannot be overlooked. We do not know if participants would adopt the technology as they say they would, at least in the proportions we found in our panel. This is a limitation of our method and data that can only be overcome with additional research. We can, however, formulate a strong working hypothesis, that, keeping all the rest constant (including the currently accepted social norms), users that have logging tools will sometimes use it offensively.

\section{Conclusion}
Throughout this paper, we have tried to convey a three-part argument, that 1) users have concerns about physical intrusion by socially close adversaries that aren't being addressed by current security technology; 2) that they can turn to activity loggers, which we show are feasible enough that someone could make them available to the general public; and 3) that such technologies could alleviate current anxieties to a certain degree, but that they could also afford kinds of practices which are themselves worrisome. We hope this work can foster what we believe to be a much needed discussion on the latter point. While we can certainly envision scenarios where people would share their devices offensively, for instance to record an unwitting friend's password, we can also see a future where no one would dare to do anything sensitive in a device that isn't their own. Indeed, if that were to become the social norm, current concerns would no longer be justifiable.

\bibliographystyle{splncs04}
\bibliography{refs}

\end{document}